\documentclass[11pt]{article}
\usepackage{amsmath}
\usepackage{amssymb}
\usepackage{graphicx,psfrag,epsf}
\usepackage{enumerate}
\usepackage{fullpage}
\usepackage{natbib}
\usepackage[usenames,dvipsnames]{xcolor}
\usepackage{url} 
\usepackage{rotating}
\usepackage{tikz}
\usepackage{picture}
\usetikzlibrary{shapes}
\usetikzlibrary{plotmarks}
\usepackage{floatrow}

\newcommand{\blind}{1}

\newtheorem{exmp}{Example}
\newcommand{\e}[1]{{\mathbb E}\left[#1 \right]}

\newcommand{\tr}{\mathop{\mathit{Tr}}\nolimits}
\DeclareMathOperator*{\argmax}{arg\,max}

\begin{document}

\def\spacingset#1{\renewcommand{\baselinestretch}%
{#1}\small\normalsize} \spacingset{1}

\if1\blind
{
  \title{\bf Optimal block designs for experiments on networks}
  \author{
	\large Vasiliki Koutra$^\dagger$\footnote{Contact: Vasiliki Koutra; \texttt{vasiliki.koutra@kcl.ac.uk}; Department of Mathematics,
King's College London, London WC2R 2LS, UK}, Steven G. Gilmour$^\dagger$, Ben M. Parker$^\ddagger$ \\[1ex]
    $^\dagger$ Department of Mathematics, King's College London \\
    $^\ddagger$ School of Computing and Engineering, University of West London \large}
 \date{\vspace{-1.2cm}}
  \maketitle
} \fi

\bigskip
\begin{abstract} 
We propose a method for constructing optimal block designs for experiments on networks. The response model for a given network interference structure extends the linear network effects model to incorporate blocks. The optimality criteria are chosen to reflect the experimental objectives and an exchange algorithm is used to search across the design space for obtaining an efficient design when an exhaustive search is not possible. Our interest lies in estimating the direct comparisons among treatments, in the presence of nuisance network effects that stem from the underlying network interference structure governing the experimental units, or in the network effects themselves. Comparisons of optimal designs under different models, including the standard treatment models, are examined by comparing the variance and bias of treatment effect estimators. We also suggest a way of defining blocks, while taking into account the interrelations of groups of experimental units within a network, using spectral clustering techniques to achieve optimal modularity. We expect connected units within closed-form communities to behave similarly to an external stimulus. We provide evidence that our approach can lead to efficiency gains over conventional designs such as randomized designs that ignore the network structure and we illustrate its usefulness for experiments on networks.
\end{abstract}


\vfill
\newpage
\section{Introduction}
\label{sec:intro}

Designing experiments on networks is a growing area of research mainly due to the rise and popularity of online social networks and viral marketing. The work presented here is motivated by the need to develop a practical methodology for obtaining efficient designs which control for variation among the experimental units from two sources: blocks and network interference, so that the true effects of the treatments can be detected. 

Consider a commercial experiment on a social networking site to compare the effectiveness of different advertisements concerning a product. The network members are connected via virtual friendships. The responses can be the quantities purchased during the week immediately following the advertising campaign. The company's goal is to maximize the appeal of the product by using the advertisement as a tool to affect the purchasing decisions of potential customers and the experiment will be used as a means for comparing different advertisements. The advertisements used might have an effect not only on the recipient but also on their (virtual) friends. Moreover, there may be blocking structures related to different age groups of customers or cliques of close friends who engage in similar behaviours, e.g.\ making similar decisions as to what quantity of product to purchase. We might expect subjects in the same block to have similar responses and subjects in different (non-overlapping) blocks to have dissimilar responses, irrespective of the presence or absence of viral effects of specific advertisements. By allowing for this in the design, we ensure more precise comparisons of the effects of advertisements. 

\begin{figure}[h!]
\begin{center}
\includegraphics[width=0.4\textwidth]{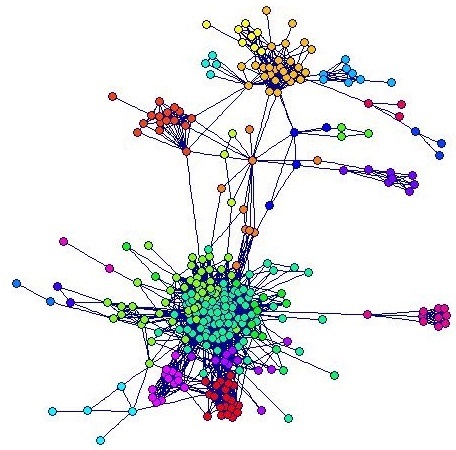} 
\end{center}
\caption{A Facebook ego-network with 24 non-overlapping blocks indicated by colours. \label{fig:FB}}
\end{figure}

In this paper we consider methodology appropriate for small to moderate sizes of networks. Applications of such networks are plentiful and examples include agriculture, biology, engineering, marketing, pharmaceuticals and other areas. As motivation, consider the network illustrated in  Figure~\ref{fig:FB}, which is a small subset of the Facebook network as obtained from the Stanford network data set collection (\url{snap.stanford.edu/data/egonets-Facebook.html}), comprising $324$ vertices (Facebook members) and $2514$ undirected edges (mutual virtual friendships) forming an ego-network. The `centre' vertex of an ego-network (the `ego') is not included in the graph, which consists of only the ego's friends (its contacts). 

A noteworthy work which accommodated interference is \cite{BesagandKempton1986}, which discussed spatial techniques and provided appropriate models to adjust for response or treatment interference in the context of agricultural field experiments. With our interest in the case of treatment interference, we focus on the model of \cite{Pearce1957} presented by \cite{BesagandKempton1986}, which jointly takes into account `local' (i.e.\ direct) and `remote' (i.e.\ indirect or neighbour) effects in the model. A wide variety of candidate models has been suggested for the case of treatment interference thereafter. Important examples include the work of \cite{Druilhet1999}, \cite{KunertandMartin2000} and \cite{KunertandMersmann2011}, who provided models and efficient designs that concern experiments where units are arranged in a circle or a line and in which neighbours one unit apart can interfere with each other. More recently, \cite{Parkeretal2016} adopted the conceptual approach of \cite{Pearce1957} and introduced a model, called the linear network effects model (LNM). The LNM is similar to the model of \cite{KunertandMartin2000}. However, it differs by relaxing the assumption of neighbour effects existing in only one direction and allows for a network setting in which units can be linked in a way that does not form a regular layout. Here, we extend the LNM of \cite{Parkeretal2016} to include the notion of blocks. 

After we introduce the network block model (NBM) (Section \ref{sec:model}), we define two specific design criteria for estimating with minimum variance the direct treatment comparisons in the presence of nuisance network effects or the network effects themselves (Section \ref{sec:design}). We describe a systematic exchange algorithm used to obtain our designs (Section \ref{sec:algorithm}). We then make some comparisons among optimal designs under different models and investigate issues associated with the design efficiency and bias arising in the analysis of dependent data (Section \ref{sec:comparisons}). When blocks arise as cliques in the network, we suggest a method to define the blocks using spectral clustering and the concept of modularity. We give an example illustrating the recommended step-by-step methodology for the special case of two unstructured treatments (Section \ref{sec:spectral_clustering}). Finally, we discuss some practical concerns emerging from the suggested methodology and further issues of interest (Section \ref{sec:disc}).

\section{Statistical model}
\label{sec:model}

Suppose that $n$ experimental units are available for experimentation and that they form a network which is represented by means of a graph $\mathcal{G}=(\mathcal{V},\mathcal{E})$, with vertex set $\mathcal{V}$ (of size $n$) and edge set $\mathcal{E}$ (of size $l$). The vertices might represent individuals connected through edges which might indicate some form of connection such as friendship, collaboration, or communication. The \textit{adjacency matrix} of a graph is an $n\times n$ matrix $A=\left[A_{jh}\right]$ with $j,h \in V$, which is a compact way to represent the collection of edges. We focus on connected, undirected and unweighted graphs with $A_{jh}=A_{hj}  \in \left\{0, 1\right\}$ representing the absence or presence of an edge between the vertices $j$ and $h$ (mutual connections). Thus $A$ is symmetric and binary. The diagonal elements represent the self-links, which by convention are set to zero. In an undirected network, the \textit{degree}, $d_j$, of the vertex $v_j \in \mathcal{V}$ denotes the number of vertices that vertex $v_j$ is connected to (number of neighbours), i.e.\ $d_{j}=\sum_{h=1}^n A_{jh}$\,.

We assume that there are groups of experimental units which are expected to give similar responses, e.g.\ because they share properties. We label these groups $1,\ldots,\kappa$ and let group $g$ have $n_{(g)}$ experimental units within it. We model the responses of experimental units by the Network Block Model (NBM), which is an extension of the LNM \citep{Parkeretal2016} and is described by the equation
\[
\mbox{NBM:} \; y_{ij} =\mu+\tau_{r(ij)}+b_i+ \sum \limits_{g=1}^{\kappa} \sum \limits_{h=1}^{n_{(g)}} A_{\{ij,gh\}} \gamma_{r(gh)}+\epsilon_{ij}\,,
\]
where $i=1,2,\ldots,\kappa; j=1,2,\ldots,n_{(i)}$, $y_{ij}$ is the response from unit $j$ in the $i$th block receiving the treatment $s=r(ij)\in \left\{1,\ldots, m\right\}$, $\mu$ represents the response for a baseline treatment or (unit) average, $b_i$ is the effect of block $i$, $\tau_{r(ij)}$ is the (direct) treatment effect, $A_{\{ij,gh\}}$/ A=$A_{\{ij,gh\}}$ with $j,h \in V$ is the adjacency matrix indicating the edge between units $j$ and $h$ belonging to blocks $i$ and $g$ $\in \left\{1,2,\ldots,\kappa \right\}$ respectively, $\gamma_{r(gh)}$ is the network effect (neighbour or indirect treatment effect) and $\epsilon_{ij}$ are the errors, which we assume to be independent and identically distributed with mean $0$ and constant variance $\sigma^2$\,. 

To overcome the model overparametrisation requires imposing some constraints, otherwise the normal equations have an infinite number of solutions and our parameters cannot be uniquely estimated. In our examples, without loss of generality, we assume the set-last-to-zero linear constraints with the $m$th treatment effect $\tau_m$ and $\kappa$th block effect $b_\kappa$ to be set equal to zero. A wider set of network structures can be tackled by by introducing a constraint on  the model parameters corresponding to the network effects, i.e. $\gamma_m=0$. For example, in practical terms that would allow regular graphs to be considered.

The expectation of the response for this model in matrix form is
\[
\e{\boldsymbol{y}} = \left(\textbf{1} \; \boldsymbol{u_1} \ldots \boldsymbol{u_{m-1}} \;  \boldsymbol{w_1} \ldots \boldsymbol{w_{\kappa-1}} \; A\boldsymbol{u_1} \ldots A\boldsymbol{u_m} \right) \boldsymbol{\beta}\,,
\]
where $\boldsymbol{\beta}={(\mu \quad \boldsymbol{\tau}^T \quad \boldsymbol{b}^T \quad \boldsymbol{\gamma}^T)}^T={(\mu \quad \tau_1\ldots \tau_{m-1} \quad b_1 \; \ldots \; b_{\kappa-1} \quad \gamma_1\ldots\gamma_m)}^T$ is the vector of parameters. There are no columns corresponding to the $m$th treatment effect and the $\kappa$th block effect since we have assumed them to be zero. Following \citet[Sec.~2.7]{Bailey2008}, vector $\textbf{u}_s$ corresponds to an $n \times 1$ vector where for each of the treatments ($s=1,2,\ldots,m$) all the elements of the vector equal zero, except for those which correspond to the units which receive that treatment and are equal to one. For instance $\textbf{u}_1$ is the indicator vector with ones for the unit(s) receiving treatment $s=1$ and zeros elsewhere. The same holds for the vector $\textbf{w}_i$ for block $i$. The symmetric information matrix $M$ is 
\[
\arraycolsep=2pt\def\arraystretch{1}
M=\left(\begin{array}{cccc}
n & \textbf{1}^T {X_\tau}^{\star}   & \textbf{1}^T{X_b}^{\star} &\textbf{1}^T AX_\tau  \\
{{X_\tau}^{\star}}^T\textbf{1} & {{X_\tau}^{\star}}^T{X_\tau}^{\star} & {{X_\tau}^{\star}}^T{X_b}^{\star} & {{X_\tau}^{\star}}^T AX_\tau  \\
{{X_b}^{\star}}^T\textbf{1} & {{X_b}^{\star}}^T{{X_\tau}^{\star}} & {{X_b}^{\star}}^T{X_b}^{\star} & {{X_b}^{\star}}^T AX_\tau  \\
{X_\tau}^TA\textbf{1} & {X_\tau}^TA{X_\tau}^{\star} &{X_\tau}^TA{X_b}^{\star} & {X_\tau}^T A^2X_\tau \end{array} \right) \,, \\[1ex]
\]
where $X_\tau$ and $X_b$ are the incidence matrices for treatment and block effects respectively and $X_\tau^{\star}$ and $X_b^{\star}$ are the same matrices with their last columns deleted. We should note at this point that only certain types of network structures will enable us to estimate the network effects. With regular graphs the estimation of network effects is not possible.

\section{Designs with block and network effects}
\label{sec:design}

In order to determine the optimal block designs we use the the \textit{L}-optimality criterion that minimizes the average variance of the estimators of a pre-specified set of linear functions of the parameters $S^T\boldsymbol{\beta}$, where $S$ corresponds to multiple vectors of known constants \citep[Ch.~10]{Atkinson2007}. The optimality criterion function to be minimised is thus a scalar function of $var(S^T \hat{\boldsymbol{\beta}})$ which is proportional to $S^T \left\{M(\xi)\right\}^{-1}S$, where $M(\xi)=M$ corresponds to the information matrix and $\xi$ is a design chosen from $\Xi$ the set of all possible designs, where the design is a choice of treatment assignments to the experimental units that correspond to the vertices of the network. The minimisation of the optimality criterion function

\[
\phi(\xi)=\tr\left(S^T \left\{M(\xi)\right\}^{-1} S\right)=\tr\left(\left\{M(\xi)\right\}^{-1}L\right)\,,
\]
where $L=SS^T$, leads to an \textit{L}-optimal design. For the \textit{L}-optimal design $\xi^*$, $\phi^* = \phi(\xi^*)=\min_{\xi \in \Xi}\phi(\xi)$, is the \textit{optimal function value}. 

To obtain these pairwise differences we have to explicitly define the $\boldsymbol{s}$ vectors that compose S. \bigskip

\noindent \textbf{Definition.} Let $s(\alpha_1,\alpha_2)$ be a contrast vector formed in the following way:
\begin{enumerate}
 \item Form a vector of $2m+\kappa+1$ zeroes, corresponding to the intercept, $m$ treatment effects, $\boldsymbol{\tau}$, followed by $\kappa$ block effects, $\mathbf{b}$, followed by $m$ network effects, $\boldsymbol{\gamma}$.
 \item Let the $\alpha_1$th and $\alpha_2$th elements be 1 and -1 respectively, corresponding to the particular effects we wish to estimate in a given contrast.
 \item Delete the $(m+1)$th and $(m+\kappa+1)$th elements, corresponding to our restriction that $\tau_m=0$ and $b_\kappa=0$ to estimate the treatment and block effects uniquely.
\end{enumerate}

The resulting vectors of length $(2m+\kappa-1)$  are pre- and post-multiplied by the $(2m+\kappa-1)\times (2m+\kappa-1)$ matrix $M^{-1}$ in the summation that defines the optimality criteria $\phi_1$ and $\phi_2$ which we now define.  

We seek to minimise the average variance of all pairwise differences of treatment effects 
\[
\frac{2}{m(m-1)} \sum_{s=1}^{m-1}\sum_{s'=s+1}^m var\left(\widehat{\tau_s-\tau_{s'}}\right)\,,\nonumber
\]
which is proportional to 
\[
\phi_1= \sum_{v=2}^{m} \sum_{q=v+1}^{m+1}\boldsymbol{s}^T(v,q)M^{-1}\boldsymbol{s}(v,q)\,. \nonumber
\]
Alternatively we seek to minimise the average variance of all pairwise differences of network effects 
\[
\phi_2=\sum_{v=m+\kappa+2}^{2m+\kappa} \sum_{q=v+1}^{2m+\kappa+1}{\boldsymbol{s}^T(v,q)M^{-1}\boldsymbol{s}(v,q)}\,.
\]
For instance, for the case of $m=2$ treatments and $\kappa=2$ blocks the criteria become 
\begin{align*}
\phi_1&=\boldsymbol{s}^T(2,3) \; M^{-1} \; \boldsymbol{s}(2,3) \\
\phi_2&=\boldsymbol{s}^T(6,7) \; M^{-1} \; \boldsymbol{s}(6,7),
\end{align*}
where $\boldsymbol{s}\left(2,3\right) = {\left(0 \quad 1 \quad 0 \quad 0 \quad 0 \right)}^T$ and $\boldsymbol{s}\left(6,7\right) = {\left( 0 \quad 0 \quad 0 \quad 1 \quad -1 \right)}^T$. The first corresponds to the estimate of the treatment effect difference, and the second to the estimate of the difference between network effects. 

We will often compare the performance of two designs using their \textit{relative efficiency}, which, with respect to the objective function $\phi$ of a design $\xi_2$ compared with a design $\xi_1$, is given by $\mbox{Eff}(\xi_1,\xi_2)=\phi(\xi_1)/\phi(\xi_2)$. We can also define the \textit{L}-\textit{efficiency} of a design $\xi$ as $\mbox{Eff}(\xi)=\mbox{Eff}(\xi^*,\xi)$, where $\xi^*$ is the \textit{L}-optimal design. We return to these definitions when assessing the performance of different optimal designs based on different models for a number of given networks.

\section{Exchange algorithm} 
\label{sec:algorithm}

When exhaustive search to find an optimal design is not possible due to the large number of units and/or treatments, we implement an approximate method. In the literature, there exist a number of computationally efficient algorithms for finding near-optimal designs in a practical amount of time using iterative methods, e.g., \cite{Fedorov1972} and \cite{CookandNachtsheim1980}. The main steps involved in the majority of those algorithms are the following: (i) initialisation of the search, e.g.\ random generation of a non-singular design (i.e.\ the matrix $M$ must be non-singular to ensure that the parameters in $\phi$ are estimable); (ii) modification of the current solution, e.g.\ make exchanges in the treatment set; (iii) assessment of new solution, i.e.\ design is assessed with respect to an objective function. Steps (ii)-(iii) are repeated until no change improves the design value; (iv) termination of the search process and return of the final design (which is assumed to be optimal). 

We adapt the `Modified Fedorov Exchange Algorithm' of \cite{CookandNachtsheim1980} to obtain an exchange algorithm for finding near-optimal designs for unstructured treatments on networks. The algorithm can be described as follows:

\bigskip
\textbf{Point Exchange on Networks (PEN) Algorithm} 
\begin{itemize}
\item[--] Step $1$: Generate a random initial design (i.e.\ random assignment of a treatment label to each unit). Units are labeled from $1$ to $n$.
\item[--] Step $2$: Calculate the optimality criterion function $\phi$ ($\phi_1$, $\phi_2$ or any other defined criterion) for the arbitrary design of step $1$.
\item[--] Step $3$: Iterate from $j=1$ to $n$ to find a better design (where $j$ corresponds to the $j$-th unit). In the $j$th iteration:
\begin{enumerate}
\item[a.] Exchange the treatment applied to unit $j$ with another treatment.  
\item[b.] Calculate the information matrix. If it is non-singular, compute the new value of the chosen criterion. Otherwise go to d. 
\item[c.] If the new criterion function value is better than the current one, keep the exchange of treatments. Otherwise, undo the exchange. 
\item[d.] If $j<n$, set $j\leftarrow j+1$.
\end{enumerate}
\item[--] Step $4$: Repeat step $3$ until a complete pass of all the vertices yields no further improvement and then go to step $5$.	
\item[--] Step $5$: Rerun all the above steps for several randomly generated initial designs and return the design with the lowest criterion function value $\phi^*$. This is the near-optimal design, which will usually be globally optimal, though this cannot be guaranteed.
End.
\end{itemize}

The last step is required to obtain an efficient final design (and overcome the problem of becoming stuck in a local optimum) by using multiple random initialisations. For the particular problems of interest in this work, the PEN algorithm appears to be powerful enough and simple to implement. 

\section{Comparison of optimal designs under different models}
\label{sec:comparisons}

We aim to construct experimental designs on networks that are efficient for estimating treatment or network effects. However, we want to investigate the bias of the estimated model parameters introduced by possible model misspecification. In this section we provide comparisons of optimal designs for estimating the treatment and network effects under traditional models that ignore network interference and models that account for it. In doing so we compare the optimal designs with network effects to the randomised balanced designs (equal replication), providing evidence that depending on the outcome of the randomisation, the balanced design is typically not very good. In the majority of cases designs that account for the network structure have higher efficiency than the standard designs. 

We also obtain the bias introduced in the parameter estimators due to the model misspecification as a function of the unknown model parameters. The models we consider are the model derived from the completely randomised design $\left\{\mbox{CRM}\right\}$, the model derived from the generalised randomised block design $\left\{\mbox{RBM}\right\}$ (which allows blocks potentially to be different sizes), the linear network effects model $\left\{\mbox{LNM}\right\}$ and the linear network with blocks model $\left\{\mbox{NBM}\right\}$, defined as:
\begin{align*}
\mbox{CRM:} \; y_{j} &= \mu + \tau_{r(j)}+ \epsilon_{j} \\[2ex]
\mbox{RBM:} \; y_{ij} &= \mu + \tau_{r(ij)}+ b_i + \epsilon_{ij} \\[.7ex]
\mbox{LNM:} \; y_{j} &= \mu + \tau_{r(j)}+ \sum \limits_{h=1}^{n} A_{\{j,h\}} \gamma_{r(h)}+\epsilon_{j} \\[.2ex]
\mbox{NBM:} \; y_{ij} &=\mu+\tau_{r(ij)}+b_i+ \sum \limits_{g=1}^{\kappa} \sum \limits_{h=1}^{n_{(g)}} A_{\{ij,gh\}} \gamma_{r(gh)}+\epsilon_{ij} \,.
\end{align*}

\begin{figure}
\begin{center}
\includegraphics[width=0.35\textwidth]{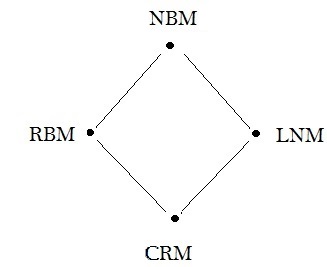} 
\end{center}
\caption{Hasse diagram for the collection of models. \label{fig:hasse}}
\end{figure}

In all cases, we assume that the errors are independent and randomly distributed with zero mean and constant variance. The CRM and RBM are the standard treatment models, that are derived from randomisation schemes without and with blocks respectively. The designs for which these models are appropriate are the simplest forms of designs to compare different treatments by randomly assigning them to experimental units (which for RBM are additionally arranged  in $\kappa$ blocks). The LNM and NBM are extensions of CRM and RBM respectively, including a network term for capturing the connections between units. Figure~\ref{fig:hasse} illustrates the hierarchy among the models by means of a Hasse diagram (see, e.g., \citealp{Bailey2008}). Here, we use the Hasse diagram to describe the collection of models considered, with dots representing models and lines representing nesting relationships between models. If the true model is a sub-model of the assumed model, the bias in estimating treatment effects will be zero. This will be further explained in the following examples. Designs which are optimal for each of these models are labelled CRD, RBD, LND and NBD respectively. 

We now compare the properties of each design for each model. In doing so we obtain the optimal function values for the two optimality criteria considering each model. Under the assumption of independent errors that have a common variance $\sigma^2$, the variance-covariance matrix of the least squares estimator $\hat{\boldsymbol{\beta}}$ is $var(\hat{\boldsymbol{\beta}}) = \sigma^2 M^{-1}$. Given that interest lies in the comparisons of the designs, the value $\sigma^2$ is not relevant since the value is the same if the model is identical for all proposed designs for a particular experiment.

\subsection{Efficiencies of randomised designs}

If we ignore the network structure in our design, we would generally assume that all experimental units in the network (for the CRD) or within a block (for the RBD) were exchangeable. There are therefore a large number of designs which we could choose as optimal under the CRD or RBD. In general, we would choose one of these equivalent designs at random, and we refer to this as a randomised design. 

In this section we compare the optimal designs with network effects to the randomised balanced designs (while ignoring the network structure), providing evidence that when under randomisation, the balanced design is typically not very good. In the majority of cases designs that account for the network structure have higher efficiency than the standard designs. The optimal randomised designs, under the CRM and the RBM, have near-equal replication of each treatment.

\begin{figure}
\begin{center}
\includegraphics[width=.4\textwidth]{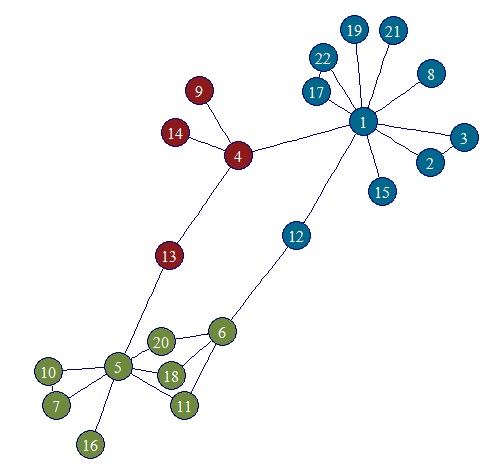} 
\end{center}
\caption{Social network with three blocks indicated by colours. \label{fig:toyexample}}
\end{figure}

\begin{exmp}\label{ex:exmp1} \end{exmp} The social network in Figure~\ref{fig:toyexample} is a co-authorship network, comprising $22$ subjects (PhD students, supervisors and co-authors) and $27$ edges, which indicate the ties between individuals with common publications, or ties between PhD students and their supervisors at the University of Southampton in the field of design of experiments in 2015. Let us assume that we want to conduct an experiment on this network in order to compare two distinct treatments. The three blocks have been defined following the method we suggest in Section \ref{sec:spectral_clustering}, but any other means of defining blocks could be addressed in the same way. 

\begin{figure}
\begin{center}
\includegraphics[width=.8\textwidth]{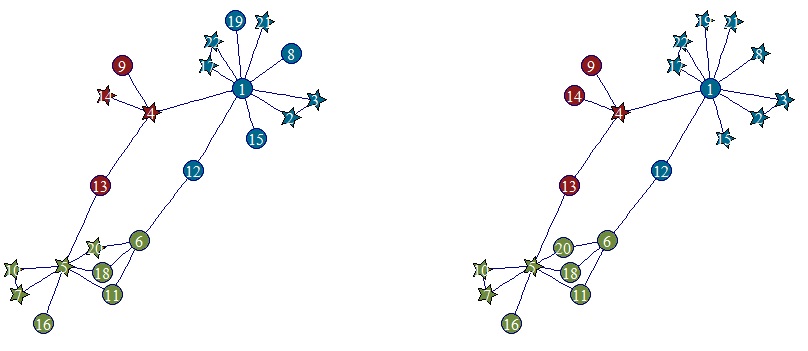} 
\end{center}
\caption{Optimal designs (under NBM), $\phi_1$ (\textit{left}) and $\phi_2$ (\textit{right}); different colours indicate different blocks and different vertex shapes different treatments. \label{fig:alloc_underLNBM}}
\end{figure}

We obtain the \textit{L}-optimal designs (using exhaustive search) for this network which are illustrated in Figure~\ref{fig:alloc_underLNBM}. Overall the optimal design for $\phi_1$ has all treatments equally allocated to the subjects, i.e.\ $11$ subjects receive each of the two treatments. Subjects allocated within each treatment have similar first and second order degrees (numbers of connections between units of distance one or two). The design for $\phi_1$ also has equal replication within each block. For $\phi_2$, the design is highly dependent on the particular network structure. For instance subjects located at the ends of the network tend to receive the same treatment which is different from their better connected immediate neighbours.

Tables \ref{tab:phi_1soton} and \ref{tab:phi_2soton} illustrate the optimality function values for each design (columns) under the different models (rows). We can obtain the efficiencies of the designs with respect to the optimal designs. The criterion function values of the optimal designs are highlighted in bold. Recall that the smaller the criterion function value the better the design is. 

A key observation derived from these tables is that a randomised design which ignores the network effects is on average highly inefficient, especially with respect to $\phi_2$. Note the values for CRD and RBD are the mean values over all $705432$ and $105840$ possible randomisations respectively. 

In Table~\ref{tab:phi_1soton}, we can observe that when assuming the LNM to be true, the NBD, which has equal replication within blocks, is almost as efficient as the LND (with efficiency $0.1822/0.1827=99.7\%$). Under NBM the LND, which has equal replication overall but not within blocks, is $50.4\%$ efficient. The poor performance of the LND relative to the NBD under the NBM results from having a completely different structure and different allocation to unequal and relatively small-sized blocks. NBD is balanced overall but also balanced within blocks. On the other hand, the optimisation process for the LNM drives the design away from having an equal treatment allocation within blocks, resulting in many neighbouring vertices receiving the same treatment. Moreover, under the NBM, a RBD which has equal replication within blocks performs better than LND on average. This means that, if we account for the blocks, we almost certainly do better. Another interesting observation is that CRD performs better than LND on average under NBM. This is because we consider all possible randomisations implying that in general randomisation will be beneficial if blocks are not known or included. It is worth noting that the optimal function values, which are $0.1822$ (under the LNM) and $0.1828$ (under the NBM) are approximately equal to the minimum average variance possible for the unstructured case when having independent units (i.e.\ $\sigma^2/n_1+\sigma^2/n_2 =2\sigma^2 m/n = 0.1818 \sigma^2$ with $n_1=n_2=n/2$).

In Table~\ref{tab:phi_2soton} the CRD and RBD perform poorly under both network models with their efficiencies being on average $16.4\%$ and $16.5\%$ under the LNM and NBM respectively when ignoring blocks (for CRD), and $12.3\%$ and $15.5\%$ under the LNM and NBM respectively when accounting for blocks (for RBD). These low efficiencies can be explained by the fact that these standard designs ignore the indirect treatment effects resulting from the network structure among units. We are able to estimate the network effects only under the LNM and the NBM. Moreover, when assuming the NBM to be true, the LND is only $39\%$ efficient.

\begin{table}
\caption{Comparisons of the designs for $\phi_1$ under different models.  \label{tab:phi_1soton}}
\begin{center}
\begin{tabular}{ccccc}
\textbf{Models} & \multicolumn{4}{c}{\textbf{Optimal designs}} \\ 
                & \textbf{CRD}       	  &\textbf{RBD}  & \textbf{LND}               & \textbf{NBD} \\ 
\textbf{CRM}    & \textbf{0.1818}	  & 0.1818    & 0.1818                  & 0.1818     \\ 
\textbf{RBM}    & 0.2034             &  \textbf{0.1818} &0.2685 						&0.1818 \\ 
\textbf{LNM}    & 0.2126						 & 0.2191     &\textbf{0.1822}            &0.1827     \\ 
\textbf{NBM}   & 0.2500						& 0.2270      &0.3621                   &\textbf{0.1828} \\ \\
\end{tabular}
\end{center}
\end{table}

\begin{table}
\caption{Comparisons of the designs for $\phi_2$ under different models.  \label{tab:phi_2soton}}
\begin{center}
\begin{tabular}{ccccc}
\textbf{Models} & \multicolumn{4}{c}{\textbf{Optimal designs}}                        \\ 
						    & \textbf{CRD}     &  \textbf{RBD}         & \textbf{LND}        & \textbf{NBD} \\ 
\textbf{LNM}    & 0.1447 &0.1927 &\textbf{0.0237} &0.0366 \\
\textbf{NBM}   & 0.2354 &0.2503 &0.0998 &\textbf{0.0388} \\ \\
\end{tabular}
\end{center}
\end{table}

\begin{figure}
\begin{center}
\includegraphics[width=0.65\textwidth]{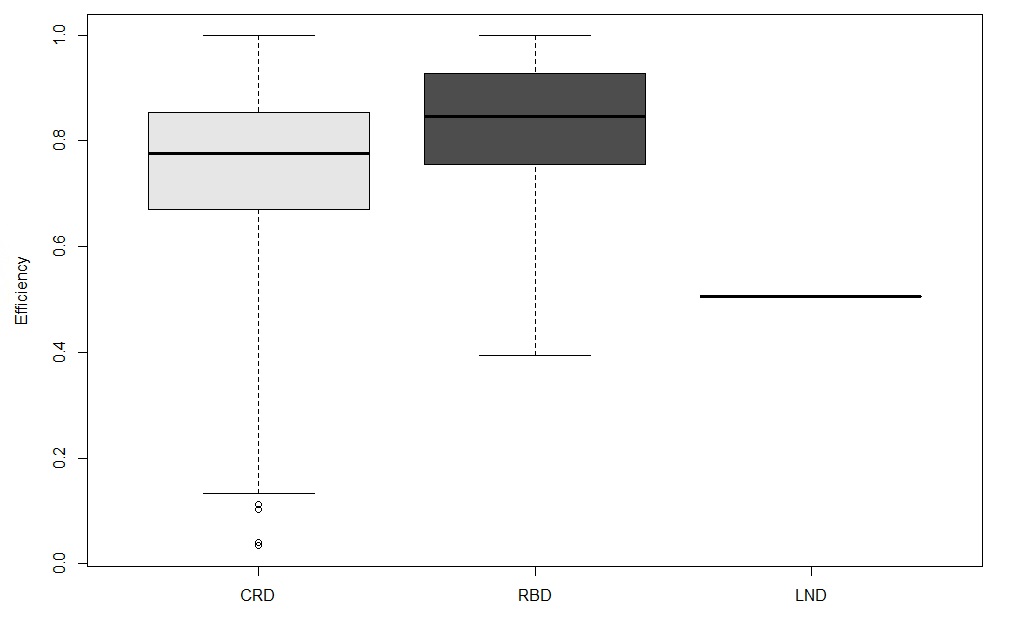} 
\end{center}
\caption{Boxplots of efficiencies for $\phi_1$ (assuming the NBM). \label{fig:boxplots_phi1_toy}}
\end{figure}

\begin{figure}
\begin{center}
\includegraphics[width=0.65\textwidth]{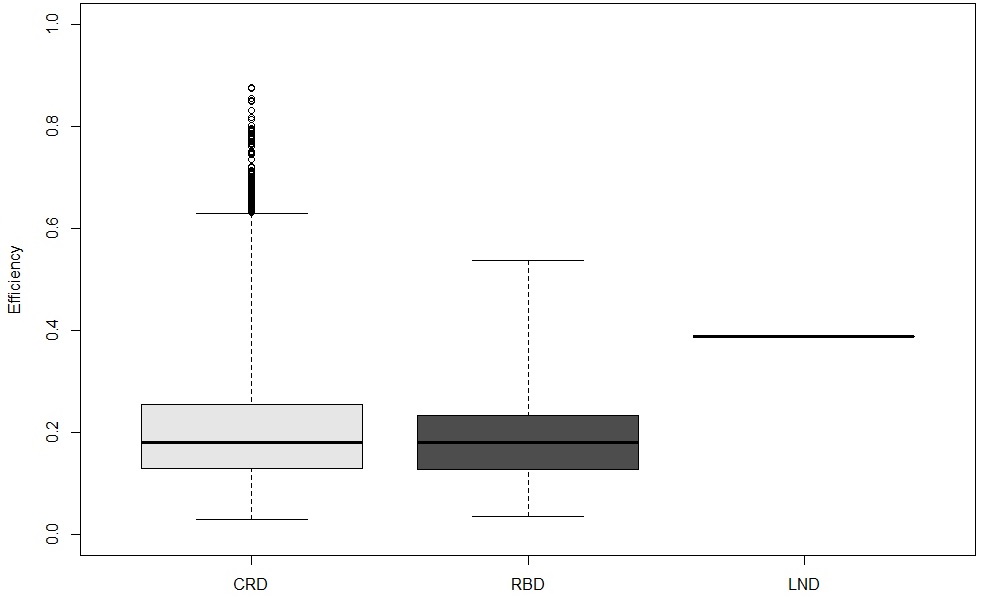} 
\end{center}
\caption{Boxplots of efficiencies for $\phi_2$ (assuming the NBM). \label{fig:boxplots_phi2_toy}}
\end{figure}

The boxplots in Figure~\ref{fig:boxplots_phi1_toy} show the \textit{L}-efficiencies of the designs based on $\phi_1$ for all CRDs and RBDs. For $\phi_1$, the CRD and RBD have median efficiencies $0.78$ and $0.83$ respectively, with minimum efficiencies $0.08$ and $0.38$ respectively. On the other hand for $\phi_2$ in Figure~\ref{fig:boxplots_phi2_toy}, CRD and RBD perform similarly with median efficiencies being $0.17$ and $0.18$, the lower quartiles being $0.05$ and $0.10$ and the upper quartiles being $0.62$ and $0.55$. This results from ignoring the network effects. The LND is approximately $50\%$ and $40\%$ as efficient as the NBD for $\phi_1$ and $\phi_2$ respectively.

\begin{exmp}\label{ex:exmp2} \end{exmp} The second example network (see Figure~\ref{fig:FB})  is a subset of the Facebook network where we wish to compare two treatments, which could correspond, for instance, to advertisements as discussed earlier. 

\begin{figure}
\begin{center}
\includegraphics[width=0.7\textwidth]{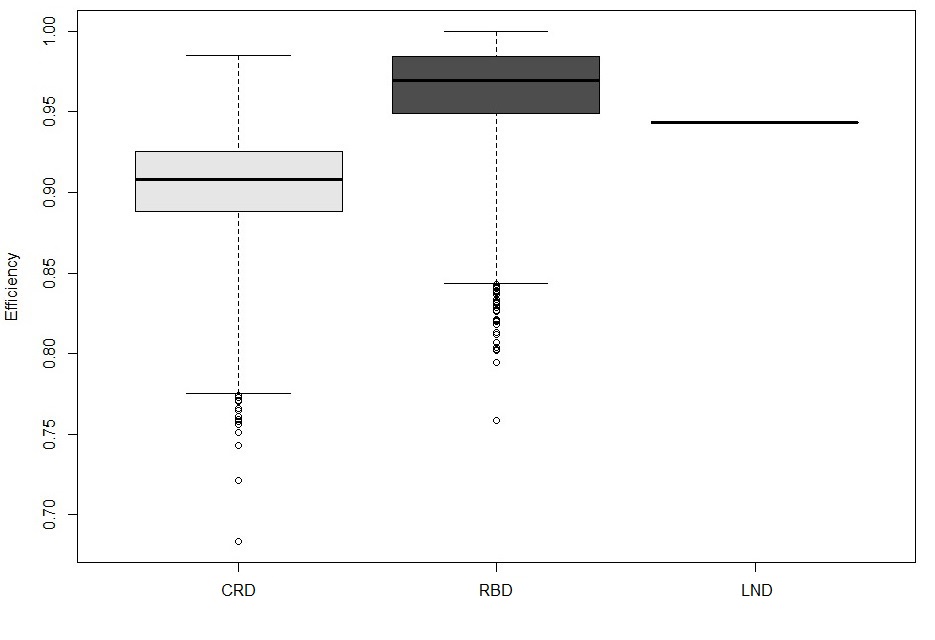} 
\end{center}
\caption{Boxplots of efficiencies for $\phi_1$. \label{fig:box1}}
\end{figure}

\begin{figure}
\begin{center}
\includegraphics[width=0.7\textwidth]{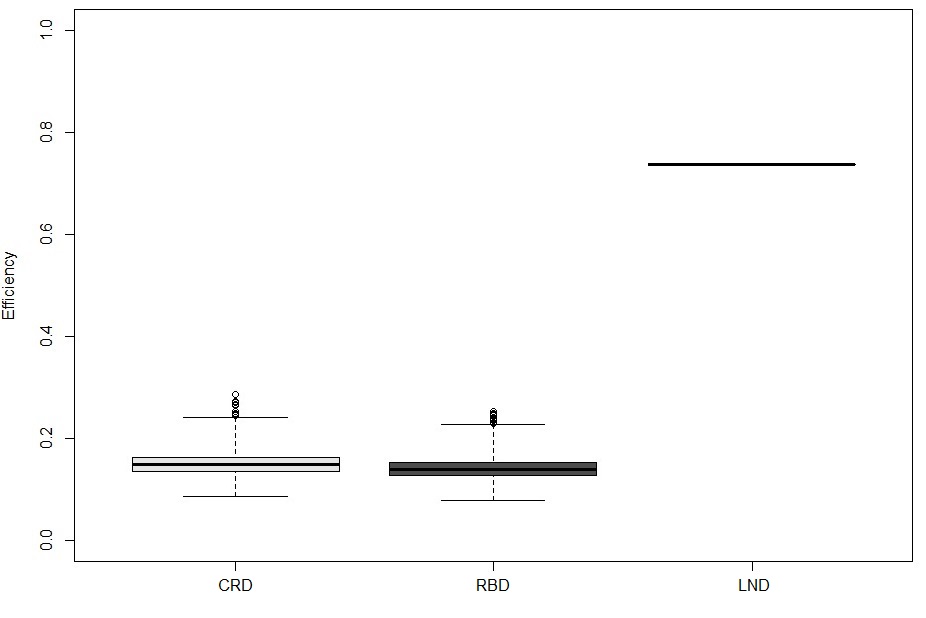}
\end{center}
\caption{Boxplots of efficiencies for $\phi_2$. \label{fig:box2}}
\end{figure}

The comparisons of the (near-)optimal designs under $\phi_1$ and under $\phi_2$ are given in Tables \ref{tab:phi1} and \ref{tab:phi2} respectively. Note the values for CRD and RBD are the mean values over a large number of possible randomisations (we have evaluated $50,000$ designs). Figures \ref{fig:box1} and \ref{fig:box2} depict the boxplots of efficiencies under \textit{L}-optimality of the design based on $\phi_1$ and $\phi_2$ respectively for a number of possible designs (chosen at random) for CRD and RBD which do not take into account network effects, although they exist. 

A key observation derived from these tables is that a randomised design, which ignores the network effects is on average highly inefficient, at least with respect to $\phi_2$. We can observe that when assuming LNM to be true, the NBD, which is completely balanced within blocks, is almost as efficient as the LND. Under NBM the LND, which is balanced overall but unbalanced with respect to blocks, is 93.9\% efficient. Moreover, under NBM, a random balanced RBD which is balanced within blocks performs better than LND on average, with the latter being as efficient as a random balanced CRD on average. In Table \ref{tab:phi2} the random balanced designs CRD and RBD perform poorly under both models
with their efficiencies ranging on average between 10.6\% and 14.8\% respectively when ignoring blocks and 8.1\% and 13.9\% respectively when accounting for blocks. These low efficiencies arise because these standard designs ignore the spillover
effects resulting from the network structure among units.

The boxplots in Figure \ref{fig:box1} depict the efficiencies under L-optimality of the design based on $\phi_1$ for a number of possible balanced designs (as chosen at random) for CRD and RBD which do not take into account network effects, although they exist. The outliers are detected by setting the upper/lower ends of the whiskers at three standard deviations. For $\phi_1$ CRD and RBD have median efficiencies 0.92 and $0.97$ respectively, with minimum efficiencies $0.68$ and $0.76$ respectively. On the other hand for $\phi_2$ in Figure \ref{fig:box2}, CRD and RBD perform similarly with median efficiencies between $0.10$ and $0.25$, lower quartiles between $0.15$ and $0.17$ and upper quartiles between $0.27$ and $0.29$. This results from not taking into account the network effects. For $\phi_1$, LND performs on average better than most of the balanced CRDs, but worse than the majority of balanced RBDs, whilst for $\phi_2$ LND is approximately $70\%$ as efficient as NBD.

\begin{table}
\caption{Comparisons of the designs for $\phi_1$ under different models.  \label{tab:phi1}}
\begin{center}
\begin{tabular}{ccccc}
\textbf{Models} & \multicolumn{4}{c}{\textbf{Optimal designs for $\phi_1 (\times 10^{2}$)}} \\ 
                & \textbf{CRD}                   & \textbf{RBD}              & \textbf{LND}                   & \textbf{NBD} \\ 
\textbf{CRM}    & \textbf{1.2346}							   & 1.2347                    & 1.2346                         & 1.2346     \\ 
\textbf{RBM}    & 1.3298                         &  \textbf{1.2432} 				 & 1.3042                         & 1.2432     \\ 
\textbf{LNM}    & 1.2481                         & 1.2747                    &  \textbf{1.2346}							  & 1.2348     \\ 
\textbf{NBM}    & 1.3749                         & 1.2907                    & 1.3177                         &  \textbf{ 1.2432} 
\end{tabular}
\end{center}
\end{table}

\begin{table}
\caption{Comparisons of the designs for $\phi_2$ under different models.  \label{tab:phi2}}
\begin{center}
\begin{tabular}{ccccc}
\textbf{Models} & \multicolumn{4}{c}{\textbf{Optimal designs for $\phi_2 (\times 10^{2}$)}}                        \\ 
						    & \textbf{CRD}  &  \textbf{RBD}       & \textbf{LND}      & \textbf{NBD} \\ 
\textbf{LNM}    & 0.1121        & 0.1474 							&	\textbf{0.0119} 	&	 0.0149 \\
\textbf{NBM}    & 0.1553        &	0.1647 							&	0.0312 						&	\textbf{0.0230} \\
\end{tabular}
\end{center}
\end{table}

\subsubsection{Recommendations}
These and other design examples led us to make the following general guidelines: 

\begin{itemize}
\item[--] On average the randomisation process assuming no network structure leads to fairly poor designs, i.e.\ high variance (this is especially true under $\phi_2$). For Example \ref{ex:exmp2}, all designs for $\phi_2$ were less than $35\%$ efficient which indicates that randomised designs are highly inefficient.
\item[--] When ignoring both block and network effects, the design can perform poorly. When taking into account the network effects but not the block effects, the design performs satisfactorily for $\phi_2$ but can perform poorly for $\phi_1$ when there is strong evidence of an underlying block structure. 
\item[--] By using blocks and ignoring network effects, with approximately a $25\%$ chance (upper quartile) one can do just as well as when taking into account the network effects, but on average one does worse (for $\phi_1$). 
\item[--] Evidence suggests that we will be better off by using blocks instead of ignoring them, independently of whether we are taking into account network effects (for $\phi_1$). 
\end{itemize}

The optimal designs under the second optimality criterion are generally unbalanced, i.e. the treatments have unequal replication in the design. Therefore, it is not surprising that the standard designs perform poorly with median efficiencies lower than $35\%$. Thus in practice, when there is a strong belief that the units are governed by a network structure and/or when blocking structure exists, then it is of importance to account for those in order to obtain an efficient design.

\subsection{Bias due to model misspecification}

In the previous section, we showed that randomised designs that ignore network effects are not efficient when used to fit the NBM. However, practitioners might use standard randomised designs and analyse their data using models that ignore network effects such as CRM and RBM. We now show that when we randomise and fit the corresponding model, the design can result in biased estimators of the quantities of interest. Although the bias has not been taken into account in the design optimality criteria, we can obtain it as a function of the model parameters and show that, under the false assumption that there are no network effects, the bias can be practically important. 

We employ Example \ref{ex:exmp1} to investigate the potential impact of model bias. We consider wrongly fitting a simpler reduced or nested model when a more complicated model is true. For example, when fitting a model with no network effect (e.g. assuming RBM) we calculate the expectation of the bias introduced in the parameter estimates for all possible balanced designs (which are optimal when there is no network effect) in terms of the true unknown parameters (for true NBM). In the Appendix, we derive the design bias in the estimators of the model parameters for the RBM, $\mathcal{W}$, due to model misspecification, i.e.\ 
\[
\mathcal{W}=\left(\begin{array}{cc} \boldsymbol{0} & B^{-1}\Gamma \\
                    \boldsymbol{0} & -I_m \end{array} \right) \boldsymbol{\beta}\,,
\]
where 
\[
B= \left(\begin{array}{ccc} \boldsymbol{1}^T \boldsymbol{1} & \boldsymbol{1}^T {X_\tau}^{\star} & \textbf{1}^T{X_b}^{\star} \\ 
{{X_\tau}^{\star}}^T\boldsymbol{1} & {{X_\tau}^{\star}}^T {X_\tau}^{\star} & {{X_\tau}^{\star}}^T {X_b}^{\star}\\
{{X_b}^{\star}}^T\textbf{1} & {{X_b}^{\star}}^T{{X_\tau}^{\star}} & {{X_b}^{\star}}^T{X_b}^{\star} \end{array}\right),
\quad \quad
\Gamma=\left(\begin{array}{c}  \boldsymbol{1}^T AX_\tau\\ 
{{X_\tau}^{\star}}^TAX_\tau\\
 {{X_b}^{\star}}^TAX_\tau \end{array}\right)\,,\\\\
\]
and $I_m$ is the $m\times m$ identity matrix.

Note that any parameters absent from the assumed model are estimated as zero, leading directly to a bias equal to the negative of the true parameter value. We obtain the biases under the different models with respect to the model parameters in Table \ref{tab:biases}. Each bias results from fitting an incorrect model and the fitted model is the same as that assumed when finding the design. Thus we can perform an experiment where we wrongly assume that there are either no network effects or no blocking effects, while the true model is of the form NBM. Note that if the true model is a submodel of the assumed model, the bias in estimating treatment effects will be zero. Moreover, the expressions resulted from averaging the biases across all possible balanced randomised designs. We see that if the network effects are substantial compared to the (direct) treatment effects, then ignoring them can potentially lead to over- or under-estimated treatment effects. Thus, by not taking into account network effects in our design, we produce an experiment which can have higher variance than necessary if the correct model is fitted and biased estimators if a simpler model is fitted.

\begin{sidewaystable}
\caption{Bias under model misspecification. \label{tab:biases}}
\begin{center}
\resizebox{\textwidth}{!}{
\begin{tabular}{ccccc}
\textbf{True model} & \multicolumn{4}{c}{\textbf{Optimal designs}}                        \\ 
			       & \textbf{CRD}       	  &\textbf{RBD}  & \textbf{LND}               & \textbf{NBD} \\ 
\textbf{CRM} & $\textbf{0}_{2\times2}$	& $\textbf{0}_{2\times 2}$ & $\textbf{0}_{2\times 2}$ & $\textbf{0}_{2\times 2}$ \\ \\ 
\textbf{RBM}    & $\left[\begin{array}{cccc}0 & 0 & 0.17 &0.34 \\0 & 0 & 0.02 &0.03 \\0 & 0 & -1 &0 \\0 & 0 & 0 &-1  \end{array} \right] \left(\begin{array}{c} \mu \\ \tau_1 \\ b_1 \\ b_2 \end{array}\right)$   
								&  $\textbf{0}_{4\times4}$ 
								&  non-nested model   
								& $\textbf{0}_{4\times4}$ \\ \\
\textbf{LNM}    & $\left[\begin{array}{cccc}0 & 0 & 1.28 &1.53 \\0 & 0 & -0.48 &-0.24 \\0 & 0 & -1 &0 \\0 & 0 & 0 &-1  \end{array} \right]\left(\begin{array}{c} \mu \\ \tau_1 \\ \gamma_1 \\ \gamma_2 \end{array}\right)$
								& non-nested model
								& $\textbf{0}_{4\times4}$           
								& $\textbf{0}_{4\times4}$      \\ \\
\textbf{NBM}   & $\left[\begin{array}{cccccc} 0 & 0 & 0.17 & 0.34 & 1.28 & 1.53\\0&0&0.02&0.03&-0.48&-0.24 \\0 & 0 & -1 & 0 & 0 & 0 \\0 & 0 & 0 &-1 & 0 & 0\\   0 & 0 & 0 & 0 &-1 & 0\\   0 & 0 & 0 & 0 & 0 &-1  \end{array}\right]\left(\begin{array}{c} \mu \\ \tau_1 \\ b_1 \\ b_2 \\ \gamma_1 \\ \gamma_2 \end{array}\right)$
								& $\left[\begin{array}{cccccc}0 & 0 & 0 &0 &1.18  &1.60\\0 & 0 & 0 &0 & -0.72 & -0.05 \\  0 & 0 & 0 & 0 & -1 & 0 \\ 0 & 0 & 0 & 0 & 0 &-1 	\\0 & 0 & 0 &0 & 0.05 & -0.45\\   0 & 0 & 0 &0 & 0.56 &-0.21 \end{array}\right]\left(\begin{array}{c} \mu \\ \tau_1 \\ b_1\\b_2\\ \gamma_1 \\ \gamma_2 \end{array}\right)$
							 & $\left[\begin{array}{cccccc}0 & 0 & 0.32 & 0.08 & 0 & 0\\0 & 0 & -0.18 & 0.52 & 0 & 0  \\ 0 & 0 & -0.01 & 0.15 & 0 & 0 \\ 0 & 0 &-0.02 &-0.13 &0 &0 \\0 & 0 & -1 & 0 & 0 & 0\\   0 & 0 & 0 &-1 & 0 & 0 \end{array}  \right]\left(\begin{array}{c} \mu \\ \tau_1 \\b_1\\b_2\\ \gamma_1 \\ \gamma_2 \end{array}\right)$
							& $\textbf{0}_{6\times6}$ 
\end{tabular}}
\end{center}
\end{sidewaystable}

We can extract the bias of the treatment effects estimates (the second line in each matrix). Assuming that we wrongly ignore network effects, we have the following two cases of the bias for the direct treatment effects estimator, where the expectation is taken over all possible randomisations
\begin{align}
\mbox{CRD under LNM:} \quad \quad \e{\hat{\tau}_1} -\tau_1&=-0.48\gamma_1-0.24\gamma_2\,, \label{eq:1}\\ 
\mbox{RBD under NBM:} \quad \quad \e{\hat{\tau}_1} -\tau_1&=-0.72\gamma_1-0.05\gamma_2\,. \label{eq:2} 
\end{align}

If the true values of the network effects are zero (so that $\gamma_1= \gamma_2=0$), the balanced design will produce unbiased estimators. We should note that the network effects are not in general of the same magnitude. However, for our investigation we assume them to be equal ($\gamma_1=\gamma_2=\gamma$), and without loss of generality we set $\gamma=1$. Under this assumption the bias in (\ref{eq:2}) is slightly larger than in (\ref{eq:1}), i.e.\ $\left|0.77\right|>\left|0.72\right|$. A conclusion drawn from this example is that if we use a model that ignores network effects there is no obvious benefit in terms of bias from including blocks. Thus having blocks does not insure us against the bias introduced by wrongly excluding network effects. 

\begin{figure}
\begin{center}
\includegraphics[width=.7\textwidth]{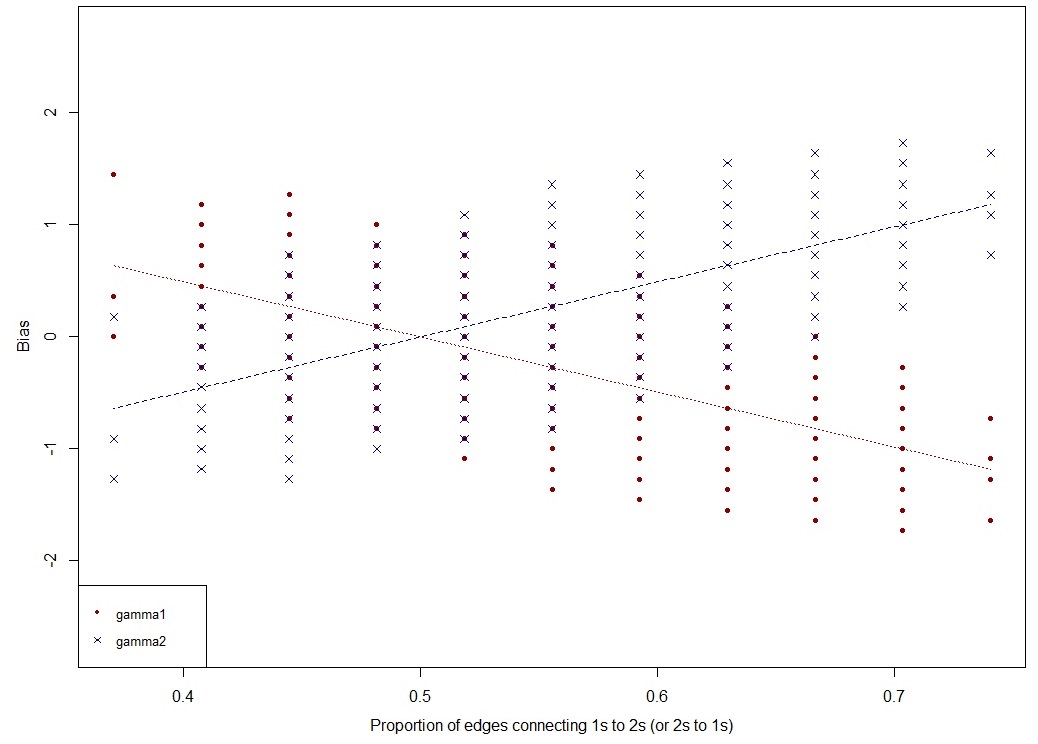} 
\end{center}
\caption{Bias in treatment effects due to network effects. \label{fig:Bias_Vs_n12}}
\end{figure}

In a completely randomised design there is a $50\%$ chance of any specific pair of units receiving both treatments $1$ and $2$. To investigate if the bias from treatment effects results from network effects passed on from units receiving treatment $1$ to those receiving treatment $2$, we focus on the number of links connecting the different treatments, which we call $l_{12}$. The plot in Figure~\ref{fig:Bias_Vs_n12} shows the bias when assuming an optimal RBD under the true NBM in treatment effect estimates, as a proportion of the true $\gamma$, against the proportion of edges, which connect pairs of units receiving different treatments. Thus the expectation of the bias is over a conditional distribution of randomisations given a fixed proportion of edges. The locations of the plotting symbols are related to the coefficients of the network effects in the bias equation under each design and are dependent on the size of the true $\gamma$. As such they correspond to the bias from RBDs for estimating the treatment effects due to the network effects under the assumption that the underlying parameters are equal to one. The  expectation of the bias over all designs for the intersection point, which corresponds to approximately half for the proportion of links, is equal to zero. It should be noted that many designs are overlapping and each location of the plotting symbols can represent the bias due to network effects for alternative RBDs. Table \ref{tab:prop} illustrates the number of balanced designs for each proportion of edges (connecting units receiving treatment 1 to units receiving treatment 2).

\begin{table}
\caption{Number of balanced designs for each proportion of edges.  \label{tab:prop}}
\begin{center}
\begin{tabular}{cccccccccccc}
\textbf{\# designs} &38  & 322 & 1532 & 4710  &9460 &12742 &11838  &7674 & 3540 &  952 &  112 \\ \hline
\textbf{prop. of edges} & 0.37  &0.41 &0.44 &0.48 &0.52 &0.56 &0.59 &0.63 &0.67 &0.7 &0.74 
\end{tabular}
\end{center}
\end{table}

\begin{figure}
\begin{center}
\includegraphics[width=.6\textwidth]{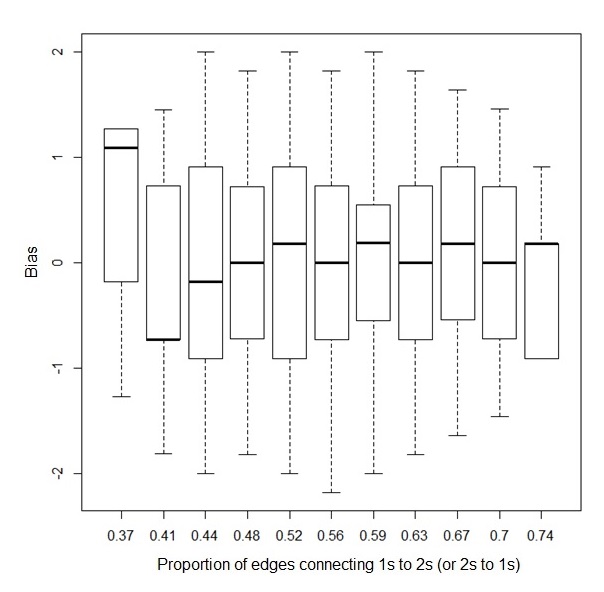} 
\end{center}
\caption{Bias for treatment effects when $\gamma_1=\gamma_2$. \label{fig:boxplots_bias}}
\end{figure}

The boxplots in Figure~\ref{fig:boxplots_bias} depict the bias in the estimation of treatment effects over all possible RBDs in the case that we assume that $\gamma_1=\gamma_2$ for every proportion of links. This expectation is conditional on the proportion of links being fixed, but averaged over the subset of randomisations which respect that. The limits of the boxplots are relative to the size of the true network effect. This plot suggests that in general the complete randomisation while ignoring the network structure does not perform well. In combination with the previous observation, we can conclude that the least expected bias in the treatment effects is achieved when the number of pairs of connected units which receive different treatments equals approximately half the total number of edges of the network (see Figure~\ref{fig:boxplots_bias} at the proportion $0.56$). 

The experimenter should impose restrictions on the allocation of treatment combinations such that the number of connected pairs of units receiving different treatments roughly equals half of the total number of edges in the network. Restrictions on the randomisation in such a way enables us to protect the experimental results against bias in treatment effects stemming from potential network effects.

\section{Defining blocks using spectral clustering}
\label{sec:spectral_clustering}

In our model, we assume that individuals within the same community exhibit similar responses, which are different to those in other communities, irrespective of the presence or absence of viral effects of specific treatment. Sometimes these communities are defined by features not directly related to the graph structure, such as age group, sex, nationality, etc. At other times they are defined by the structure of the graph itself. Then we have to deduce the community structure by using a clustering algorithm.

Spectral clustering, as the name implies, makes use of the graph spectrum, i.e.\ the set of eigenvalues and eigenvectors of a matrix describing the graph structure, to group vertices into communities. The graph spectrum plays a major role in the understanding of the structure and dynamics in the network since it is linked with numerous graph properties \citep{Chung1997,VonLuxburg2007}. Spectral clustering is a versatile clustering method which can be applied to any network and is considered to give good quality solutions. To identify the number of clusters, we additionally use the modularity of the network, which is a measure which quantifies the strength of the community structure in a network \citep{Newman2006}. 

Apart from the adjacency matrix, a graph $\mathcal{G}$ can be represented by other connectivity matrices such as the Laplacian matrices. The main differences between spectral clustering techniques lie in the choice of the Laplacian matrix. We use the normalised Laplacian graph, $L_{rw}=I-D^{-1} A$, where $D$ is the $n \times n$ diagonal degree matrix (where the entries are the degrees of the vertices). In fact, $D^{-1} A$ is the transition matrix of a standard random walk on the given graph, making it a useful tool for capturing a diffusion process, such as the treatment propagation effects, in a network.  Recall that a walk is a sequence of edges connected to a sequence of vertices, where vertices can appear more than once. A random walk is a walk across a network created by taking repeated random steps. The vertices with high degree are more likely to be visited by the random walk because there are more ways of reaching them. 

For detecting communities we implement the normalised spectral clustering algorithm, namely the Shi and Malik (SM) algorithm \citep{SM2000} as described by \cite{VonLuxburg2007}. We describe the SM algorithm in the Appendix with appropriate adjustments for the purposes of this work. This clustering algorithm gives no hint about the choice of the number of clusters or `goodness' of each partition. To quantify the quality of partitions and to choose the number of communities based on the `best' partition, we use modularity \citep{Newman2004}, which is defined as the difference between the fraction of the edges that fall within clusters and the fraction of the edges that would be expected to fall within the clusters if the edges were assigned randomly but keeping the degrees of the vertices unchanged. When the communities are not stronger than the random partition or when the network does not exhibit any community structure, the modularity score $Q$ is zero or negative (see Appendix for the detailed definition). As such $\kappa$ is regarded as the appropriate candidate for indicating the number of intrinsic communities which will define the blocks to be used in the design process, if the $\kappa$th partition corresponds to the highest modularity score over all partitions.

Revisiting Example \ref{ex:exmp2} we provide and implement the step-by-step methodology.

\begin{figure}
\begin{center}
\includegraphics[width=0.6\textwidth]{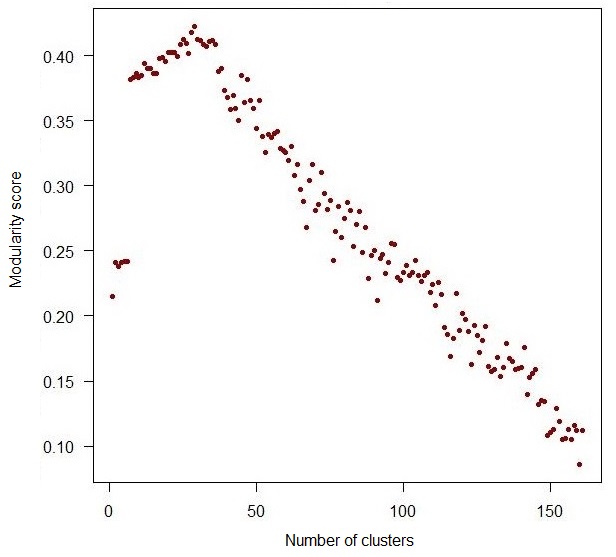} 
\end{center}
\caption{Modularity values for partitions of $\kappa=2,\ldots,162$. \label{fig:mod}}
\end{figure}

\textit{Step one: Blocking Structure}. Expressing the topology of the network through the normalised Laplacian matrix $L_{rw}$, we produce a number of possible network partitions for different values of $\kappa$ (dimensionality of the eigenvector space, i.e.\ the number of $\kappa$ eigenvectors) using the spectral clustering algorithm SM. Then we assess all the partitions produced using the quality function of modularity and choose the number of communities (fixed $\kappa$) to be used for the block design, which maximises the value of modularity. For the social network at hand the maximum modularity value over all possible partitions is found for $\kappa=24$ clusters. Figure~\ref{fig:mod} illustrates the modularity scores for different numbers of communities obtained via the graph partition. This screening stage helps us to choose the number of communities, where the modularity score takes its maximum value. The resulting clustering is illustrated in Figure~\ref{fig:FB} with different colours indicating the different $24$ blocks. 

\textit{Step two: Optimal design with block and network effects (using PEN)}. The optimal function values resulting from this search (allowing for multiple initial designs) were found to be ${\phi_1}^*=0.0124$ and ${\phi_2}^*=0.0002$. Similar to the first example, ${\phi_1}^*$ is approximately equal to the estimated variance of the difference of the two treatments when the units are independent and the model does not contain network or block effects (that is $0.0123\sigma^2$). Whilst the optimal design for estimating $\tau_1$ has $161$ and $163$ units receiving treatments $1$ and $2$ respectively, the optimal design for estimating the difference in network effects has $117$ and $207$ units receiving treatments $1$ and $2$ respectively. 

\section{Discussion}
\label{sec:disc}

Many experiments can be regarded as being on networks by appropriate specification of the adjacency matrix. The suggested methods are easily adaptable to a wide class of networked experiments and to different block definitions for enabling the researcher to make comparisons between different treatments both directly and indirectly. However, generalising these methods to large-scale networks with hundreds of thousands of nodes, for example, commercial databases, is an open research problem.

This paper offers a framework for finding optimal block designs with network effects. It provides evidence that conventional designs such as randomised designs while ignoring the network structure are inefficient compared to block designs with network effects. We saw that a model which wrongly ignores block and/or network effects leads to inefficient designs. 

We also suggest a way to define blocks, while taking into account the community structure. We justify our choice of using spectral clustering due to the simplicity of its implementation (standard linear algebra), without requiring any explicit distributional model and without making any kind of assumption on the formation of the clusters. We have three main components: $(i)$ spectral clustering of the given social network to project the vertices of the network onto an eigenvector-space; $(ii)$ modularity to choose the `best' partition of the network over all partitions; and $(iii)$ optimal block designs, for the clusters of subjects produced. 

Based on a wide variety of examples similar to the ones we presented, we can suggest some general guidelines. Heuristically the experimental design for estimating the treatment effects tends to have equal replication within each block on the units having a similar number of connections, while for estimating the network effects the design is greatly influenced by the network's first and second order connections.

\appendix
\section*{Appendix}
\label{appendix}

\subsection*{Bias due to model misspecification}

We provide the derivation of the bias in the estimators of the model parameters due to model misspecification. 

Consider performing an experiment on a network, where the adjacency matrix, $A$, is given. We assume that the true model is the NBM. However, the postulated model for the experiment is the RBM, which ignores network effects (i.e.\ $\boldsymbol{\gamma} = \boldsymbol{0}$). We want to obtain the bias in the parameter estimates due to the model misspecification as a function of the unknown model parameters. This will be accomplished by using generalised inverse matrices \citep[Ch.9]{Harville1997}. Let $\hat{\boldsymbol{\beta}}_C$ and $\hat{\boldsymbol{\beta}}_R$ be estimators of $\boldsymbol{\beta}$ with $X_C= {\left(\boldsymbol{1} \; {X_\tau}^{\star} \; {X_b}^{\star} \; AX_\tau\right)}^T$ and $X_R={\left(\boldsymbol 1 \; \: {X_\tau}^{\star} \; \: {X_b}^{\star} \; \boldsymbol{0}_{n\times m}\right)}^T$ being the extended design matrices under the NBM and RBM respectively. Note that $\hat{\boldsymbol{\beta}}_C$ is the best linear unbiased estimator for $\boldsymbol{\beta}$ so that $\e{\hat{\boldsymbol{\beta}}_C}=\boldsymbol{\beta}$ and $\hat{\boldsymbol{\beta}}_R$ is the OLS estimator of $\boldsymbol{\beta}$ for the RBM (without network effects). With the necessary algebraic calculations on generalised inverse matrices, it follows that the bias of the design, $\mathcal{W}$, under the assumption that there are network effects but we do not take them into account, is
\begin{align*}
\mathcal{W}&=\e {\hat{\boldsymbol{\beta}}_R-\hat{\boldsymbol{\beta}}_C}
= \e{{\left({X_R}^T X_R \right)}^{-}{X_R}^T \boldsymbol{y} - {\left({X_C}^T X_C\right)}^{-}{X_C}^T \boldsymbol{y}}\\ 
&= \left({\left({X_R}^T X_R\right)}^{-}{X_R}^T   - {\left({X_C}^T X_C\right)}^{-}{X_C}^T \right) \e{\boldsymbol{y}} \\ 
&= \left({\left({X_R}^T X_R\right)}^{-}{X_R}^T  - {\left({X_C}^T X_C\right)}^{-}{X_C}^T \right) X_C \boldsymbol{\beta}\\
&= \left({\left({X_R}^T X_R\right)}^{-}{X_R}^T X_C   - I \right) \boldsymbol{\beta}\\
&= \left({\left({\left(\boldsymbol 1 \; \: {X_\tau}^{\star} \; \: {X_b}^{\star} \; \boldsymbol{0}_{n\times m}\right)}^T 
\left(\begin{array}{c} \boldsymbol{1} \\{X_\tau}^{\star} \\ {X_b}^{\star} \\ \boldsymbol{0}_{n\times m} \end{array} \right)
 \right)}^{-}{\left(\boldsymbol 1 \; \: {X_\tau}^{\star} \; \: {X_b}^{\star}  \; \boldsymbol{0}_{n\times m}\right)}^T  \left(\begin{array}{c} \boldsymbol{1} \\ {X_\tau}^{\star} \\ {X_b}^{\star} \\ AX_\tau \end{array} \right)  - I\right)  \boldsymbol{\beta} \\[2ex]
&=\left({\left(
\begin{array}{cccc}
\boldsymbol{1}^T \boldsymbol{1} & \textbf{1}^T {X_\tau}^{\star}   & \textbf{1}^T{X_b}^{\star} & \boldsymbol{0}  \\
{{X_\tau}^{\star}}^T\textbf{1} & {{X_\tau}^{\star}}^T{X_\tau}^{\star} & {{X_\tau}^{\star}}^T{X_b}^{\star} & \boldsymbol{0}   \\
{{X_b}^{\star}}^T\textbf{1} & {{X_b}^{\star}}^T{{X_\tau}^{\star}} & {{X_b}^{\star}}^T{X_b}^{\star} & \boldsymbol{0}   \\
\boldsymbol{0}  & \boldsymbol{0}  &\boldsymbol{0}  & \boldsymbol{0}  \end{array}
 \right)}^{-}
\left(
\begin{array}{cccc} 
\boldsymbol{1}^T \boldsymbol{1} & \boldsymbol{1}^T {X_\tau}^{\star}  & \textbf{1}^T{X_b}^{\star}& \boldsymbol{1}^T AX_\tau\\ 
{{X_\tau}^{\star}}^T\boldsymbol{1} & {{X_\tau}^{\star}}^T {X_\tau}^{\star} & {{X_\tau}^{\star}}^T {X_b}^{\star} &{{X_\tau}^{\star}}^TAX_\tau \\
{{X_b}^{\star}}^T\textbf{1} & {{X_b}^{\star}}^T{{X_\tau}^{\star}} & {{X_b}^{\star}}^T{X_b}^{\star} & {{X_b}^{\star}}^T AX_\tau   \\
\boldsymbol{0} & \boldsymbol{0} & \boldsymbol{0} & \boldsymbol{0} \end{array} \right)- I
\right) \boldsymbol{\beta},
\end{align*}
where $I$ is the $(2m+\kappa-1) \times (2m+\kappa-1)$ identity matrix. Let $B$ and $\Gamma$ represent the $(m+\kappa-1) \times (m+\kappa-1)$ and $(m+\kappa-1) \times (m-1)$ matrices respectively 
\[
B= \left(\begin{array}{ccc} \boldsymbol{1}^T \boldsymbol{1} & \boldsymbol{1}^T {X_\tau}^{\star} & \textbf{1}^T{X_b}^{\star} \\ 
{{X_\tau}^{\star}}^T\boldsymbol{1} & {{X_\tau}^{\star}}^T {X_\tau}^{\star} & {{X_\tau}^{\star}}^T {X_b}^{\star}\\
{{X_b}^{\star}}^T\textbf{1} & {{X_b}^{\star}}^T{{X_\tau}^{\star}} & {{X_b}^{\star}}^T{X_b}^{\star} \end{array}\right)
\quad \mbox{and} \quad
\Gamma=\left(\begin{array}{c}  \boldsymbol{1}^T AX_\tau\\ 
{{X_\tau}^{\star}}^TAX_\tau\\
 {{X_b}^{\star}}^TAX_\tau \end{array}\right)\,.\\\\
\]
Then for the $(2m+\kappa-1) \times (2m+\kappa-1)$ block-diagonal matrix 
\[
\Delta=\left(\begin{array}{cc} B & \boldsymbol{0} \\
 \boldsymbol{0} & \boldsymbol{0} \end{array} \right)\,,\\
\]
we have that $B$ is a non-singular matrix (it has full rank) and, defining a $(2m+\kappa-1) \times (2m+\kappa-1)$ matrix 
\[
 G=\left(\begin{array}{cc} G_{11} & G_{12} \\
 G_{21} & G_{22} \end{array} \right)\\\\
\]
where $G_{11}$ is of dimension $(m+\kappa-1) \times (m+\kappa-1)$, we obtain
\[
\Delta G\Delta=\left(\begin{array}{cc} BG_{11}B & \boldsymbol{0} \\
 \boldsymbol{0} & \boldsymbol{0} \end{array} \right)\,,\\\\
\]
implying that $G$ is a generalised inverse of $\Delta$ if and only if $BG_{11}B=B$, or if and only if $G_{11}=B^{-1}$. Hence, we have
\[
\mathcal{W} =\left(\left(\begin{array}{cc} {B}^{-1} & \boldsymbol{0} \\
 \boldsymbol{0} & \boldsymbol{0} \end{array} \right)
\left(\begin{array}{cc} B & \Gamma \\
 \boldsymbol{0} & \boldsymbol{0} \end{array} \right)-I \right) \boldsymbol{\beta} 
=\left(\left(\begin{array}{cc} I & B^{-1}\Gamma \\
 \boldsymbol{0} & \boldsymbol{0} \end{array} \right)-I \right) \boldsymbol{\beta} 
=\left(\begin{array}{cc} \boldsymbol{0} & B^{-1}\Gamma \\
 \boldsymbol{0} & -I_m \end{array} \right) \boldsymbol{\beta}\,.\\\\
\]

Thus the bias introduced in the estimates of the parameters, $\hat{\boldsymbol{\beta}}_R$, under the false assumption
that there are no network effects is given by the quantity $B^{-1}\Gamma$. This quantity is the result of ignoring the network effects, which is represented by an adjustment of the intercept and the treatment effect estimates. Observe that $\e{\hat{\boldsymbol{\beta}}_R}\neq \e{\hat{\boldsymbol{\beta}}_C}$, unless $B^{-1}\Gamma=\boldsymbol{0}$ which results from $\Gamma=\boldsymbol{0}$ or $\gamma=\boldsymbol{0}$.

Following the same rationale, we can end up in a similar expression of the design bias introduced in the treatment effects, when we perform an experiment where we wrongly assume there are no blocking effects. This means that if we consider the reduced model LNM, while the true model is of the form NBM, the design matrix will be of the form $X_R={\left(\boldsymbol 1 \; \: {X_\tau}^{\star} \; \: \boldsymbol{0}_{n\times (\kappa-1)} \; \: AX_\tau \right)}^T$.

\subsection*{The Shi and Malik algorithm}

The SM algorithm involves three steps: I. \textit{Compute the normalised graph Laplacian} $L_{rw}$ and its spectrum (as based on the known adjacency matrix of the network); II. \textit{Dimensionality reduction}: using the $\kappa$ first eigenvectors $\boldsymbol{\upsilon}_1,\ldots,\boldsymbol{\upsilon}_\kappa$ of the graph Laplacian that correspond to the first $\kappa$ eigenvalues sorted in ascending order, let $U\in R_{n \times \kappa}$ be the matrix containing $\boldsymbol{\upsilon}_1,\ldots,\boldsymbol{\upsilon}_\kappa$ as columns. (Note that the eigenvector-space varies based on the chosen dimensionality $\kappa$.); III. \textit{Clustering step}: treating each row of $U$ as a data point, $(y_i)_{i=1,\ldots,n} \in R^\kappa$,  group them via the (standard) k-means algorithm into $\kappa$ (dimensionality of the eigenvector space) clusters, $C_1,\ldots,C_\kappa$. Therefore, the vertices of the network are projected into a $\kappa$-dimensional space, where $\kappa$ is the number  of the first nontrivial eigenvectors of $L_{rw}$. As a result, each unit is allocated to one cluster. 

We perform the clustering step for various numbers of clusters $\kappa$, $2\leq\kappa\leq n/2$, the upper bound chosen so that there should be at least two units within a cluster for achieving treatment comparisons. Note that the partition method can rely on a different clustering step rather than k-means if required. However, we used this standard one, which is the most used partitioning method.

This clustering algorithm gives no hint about the choice of the number of clusters or `goodness' of each partition. To quantify the quality of partitions and to choose the number of communities based on the `best' partition, we use modularity, 

\[
Q=\frac{1}{2l} \sum_{j,h=1}^n\sum_{i=1}^\kappa {\left(A_{jh}-\frac{d_j d_h}{2l} \right) s_{ji} s_{hi}}\,,
\]

\noindent where $s_{ji}$ and $s_{hi}$ are binary indicators of whether vertices $j$ and $h$ belong to group $i$ or not (membership vectors) and $2l=\sum_{j=1}^n{d_j}$ is the total degree of all the vertices. In other words, modularity is defined as the difference between the fraction of the edges that fall within clusters and the fraction of the edges that would be expected to fall within the clusters if the edges were assigned randomly but keeping the degrees of the vertices unchanged. The factor of $1/2$ accounts for the fact that every vertex pair $j, h$ is counted twice. When the communities are not better than the random partition or when the network does not exhibit any community structure, $Q$ is zero or negative. 

As such $\kappa$ is regarded as the appropriate candidate for indicating the number of intrinsic communities which will define the blocks to be used in the design process, if the $\kappa$th is the highest modularity score over all produced partitions, i.e.\ $\kappa=\argmax Q$.

\section*{Acknowledgements}
The authors gratefully acknowledge the Economic and Social Research Council (ESRC) for funding this research.

\bibliographystyle{Chicago}
\bibliography{mybib}
\end{document}